\documentclass[journal, one column, draftcls, 12pt]{IEEEtran}

\usepackage{amsmath}
\usepackage{graphicx}
\usepackage{amssymb}
\usepackage{cases}
\usepackage{color}
\usepackage{cite}
\usepackage[ruled,vlined,lined,ruled,linesnumbered]{algorithm2e}
\usepackage{relsize}
\usepackage[nodisplayskipstretch]{setspace}
\usepackage[top=1in, bottom=1in, left=0.7in, right=0.7in]{geometry}

\newcommand{\D}{\tilde{D}}
\newcommand{\upa}{\mathrm{upa}}
\newcommand{\lens}{\mathrm{lens}}
\newcommand{\rf}{\mathrm{rf}}
\newcommand{\dig}{\mathrm{dig}}
\newcommand{\hybrid}{\mathrm{hybrid}}
\newcommand{\sinc}{\mathrm{sinc}}

\graphicspath{{./figs/}}

\begin{document}
\title{Cost-Effective Millimeter Wave Communications with Lens Antenna Array}
\author{Yong~Zeng and Rui~Zhang \\
\thanks{The authors are with the Department of Electrical and Computer Engineering, National University of Singapore. e-mail: \{elezeng, elezhang\}@nus.edu.sg.}
}
\maketitle

\begin{abstract}
Millimeter wave (mmWave) communication is a promising technology for the fifth-generation (5G) wireless system. However, the large number of antennas used and the wide signal bandwidth in mmWave systems render the conventional multi-antenna techniques increasingly costly in terms of signal processing complexity, hardware implementation, and power consumption. In this article, we investigate cost-effective mmWave communications by first providing an overview of the main existing techniques that offer different trade-offs between performance and cost, and then focusing our discussion on a promising new technique based on the advanced {\it lens antenna array}. It is revealed that by exploiting the {\it angle-dependent energy focusing} property of lens arrays, together with the {\it angular sparsity} of the mmWave channels, mmWave lens-antenna system is able to achieve the capacity-optimal performance with very few radio-frequency (RF) chains and using the low-complexity single-carrier transmission, even for wide-band frequency-selective channels. Numerical results show that the lens-based system significantly outperforms the state-of-the-art designs for mmWave systems in both  spectrum efficiency and energy efficiency.
\end{abstract}

\section{Introduction}
Millimeter wave (mmWave) communication over the spectrum from around 30GHz to 300GHz has emerged as one of the key enabling technologies for the fifth-generation (5G) wireless systems \cite{569}. In fact, mmWave systems over the unlicensed 60GHz has already been standardized for indoor communications, such as wireless personal area networks (WPAN) \cite{571}. More recently, the FCC (Federal Communications Commission) of the United States has announced its spectrum plan to open up 3.85 GHz of licensed and 7 GHz of unlicensed high-frequency spectrum above 24GHz for wireless broadband communications\footnote{See the website https://www.fcc.gov/document/fcc-adopts-rules-facilitate-next-generation-wireless-technologies}, which makes a crucial step towards the practical application of mmWave technology in 5G.

MmWave communications possess several unique characteristics as compared to the lower-frequency counterparts. Firstly, owing to the fundamental Friss transmission equation, mmWave signals typically incur much higher free-space path loss given the same distance and antenna gains. This thus requires large (in terms of the number of array elements) antenna arrays to be equipped at the base station (BS), and also possibly at the mobile station (MS) to achieve highly directional transmission for reasonable signal coverage. Fortunately, thanks to the reduced signal wavelength, packing large arrays compactly with small form factors is feasible at mmWave frequencies, and the manufacturing of antenna arrays  itself is becoming less expensive \cite{569}. Secondly, due to the poor scattering and diffraction capabilities for high-frequency signals, mmWave channels typically have only limited number of multi-paths \cite{569}. Correspondingly, the number of angular directions that the transmit and receive beamformers need to deal with is small, which we refer to as {\it angular sparsity} for mmWave channels. 
 Thirdly, mmWave systems are expected to operate over rather large bandwidth, as the main motivation for shifting to mmWave frequencies. For instance, the current WPAN mmWave systems have allocated channels of 2.16GHz bandwidth, which is much wider than the 20MHz bandwidth in current 4G systems. As a result, mmWave systems need to support much higher sampling rates (e.g., giga-samples per second), and are expected to operate over frequency-selective channels in general, for which the detrimental inter-symbol interference (ISI) needs to be effectively mitigated.

The unique characteristics mentioned above bring  new challenges in the design of mmWave communication systems. Firstly, the large signal dimensions, resulting  from the use of both large antenna arrays in spatial domain and high signal sampling rate in time domain, make the signal processing  complexity significantly higher  in mmWave systems than lower-frequency systems. 
Secondly, the traditional  MIMO (multiple-input multiple-output) transceiver architecture which requires one dedicated radio-frequency (RF) chain for each antenna is  also more costly  in terms of both hardware implementation and power consumption, since the main  RF chain components such as local oscillators, mixers, amplifiers, and analog-to-digital converters (ADCs) all need to operate in mmWave systems at much higher frequency and over significantly larger bandwidth than their low-frequency counterparts. Thus, innovative solutions need to be devised to achieve the performance benefits of mmWave large MIMO systems, but with reduced hardware and power consumption costs. Last but not least, ISI mitigation is a non-trivial task for wide-band mmWave large MIMO systems, especially for practical designs that need to operate with low signal processing, hardware, and power consumption costs, thus rendering the traditional ISI mitigation techniques such as MIMO-OFDM (orthogonal frequency division multiplexing) or sophisticated time/frequency-domain  equalizations less appealing.

This article aims to investigate cost-effective techniques for mmWave communication. First, we provide a brief overview of the main existing techniques that offer various trade-offs between performance and cost for mmWave systems. Then we will focus our discussion on a promising new technique based on the advanced {\it lens antenna arrays}. It is shown that by exploiting the unique {\it angle-dependent energy focusing} capability of lens arrays, together with the {\it angular sparsity} of mmWave channels, mmWave lens MIMO sytem is able to achieve the capacity-optimal performance and yet overcome the major challenges discussed above, including significantly reducing the signal processing, hardware, and power consumption costs, as well as achieving inherent ISI mitigation using the low-complexity single-carrier (SC) transmission without the need of sophisticated equalization. 

\section{Cost-Aware mmWave Communication: Existing Solutions}
  Many research efforts have been devoted to reducing the RF chain costs in mmWave communications  based on two main approaches: using fewer RF chains 
   or using simpler RF chain components, as will be reviewed in this section. 
\subsection{Antenna Selection}
Antenna selection is a low-complexity scheme to reduce the number of required RF chains while still capturing the many benefits of multi-antenna systems \cite{370}. For a transceiver with $M$ antennas,  only a subset of $M_\rf<M$ antennas are selected to be activate each time and thus the number of required RF chains is reduced from $M$ to $M_\rf$. For independent fading channels, antenna selection retains the diversity degree as compared to the full-MIMO system \cite{370}. 
 Besides, when applied for spatial multiplexing, antenna selection results in comparable rate performance as the full-MIMO system, provided that $M_\rf$ is as large as the number of data streams.  However, for highly correlated channels, as usually the case for mmWave systems due to the multi-path sparsity, antenna selection usually yields poor performance due to its severe degradation in both diversity gain as well as power (beamforming) gain. For instance, in the extreme case of fully correlated channels, e.g., conventional uniform planar arrays (UPAs) in solely line-of-sight (LoS) environment, selecting $M_\rf$ out of the $M$ antennas only retains a fraction  $M_\rf/M$ of the total channel power, which is insignificant if $M_\rf\ll M$. Another drawback of antenna selection is its ineffectiveness in frequency-selective systems, since the frequency-dependent channel responses usually require different antennas to be selected at different frequencies, which makes it less effective.

\subsection{Hybrid Analog/Digital Processing}
Hybrid analog/digital processing is an effective technique to achieve a flexible trade-off between performance and cost via two-stage signal processing \cite{578,827,825,832}, i.e., a low-dimensional baseband digital processing using limited number of RF chains concatenated with an RF-band analog processing through a network of phase shifters. With full-array hybrid processing for a transceiver with $M$ antennas and $M_\rf<M$ RF chains, the output/input of each of the $M_\rf$ RF chains is connected with all the $M$ antennas via power splitters/combiners and a network of analog phase shifters. It has been shown that for narrow-band systems, hybrid processing is able to achieve the same performance as the fully digital processing scheme, provided that $M_\rf$ is at least twice of the number of data streams \cite{827}. However, the hybrid architecture requires a total number of $MM_\rf$ phase shifters, which is excessive for large systems and requires extra power consumption that may even outweight the power saving due to the use of fewer RF chains. Moreover, as compared to the fully digital systems, the precoding/beamformer design as well as channel estimation are more complicated for hybrid processing schemes, especially for wide-band systems with frequency selectivity \cite{825,832}. 
In the extreme case of $M_\rf=1$, the hybrid processing scheme reduces to {\it analog beamforming}, which requires only one RF chain and $M$ phase shifters for each transceiver, but usually at the cost of severe performance degradation due to its limited design degrees of freedom as well as its inability to support spatial multiplexing.

\subsection{Low-Resolution ADC Receiver}
Besides using fewer RF chains, an alternative approach for reducing the cost of large-array systems is to employ less expensive and more power-efficient RF chain components, such as low-resolution ADC at the receiver \cite{828}. For an ADC with $b$-bit resolution (i.e., $b$ bits per sample), it in general requires $2^b-1$ comparators. As a result, the ADC power consumption increases exponentially with its resolution. This thus provides a practical justification for using low-resolution ADC to reduce the RF chain power consumption. For mmWave systems with large bandwidth and hence requiring high sampling rates, using low-resolution ADC (e.g., 1-3 bits per sample) is even more appealing as compared to the low-frequency systems. In fact, it has been shown that in low-SNR (signal-to-noise ratio) regime, using low-resolution ADC only incurs small loss in spectral efficiency as compared to ideal ADC. However, as SNR increases, its performance loss could be substantial. For instance, for a receiver with $M$ antennas each connected with two 1-bit ADCs (for in-phase and quadrature phase components, respectively), the total number of quantized outputs is maximally $2^{2M}$, and hence its maximum achievable data rate is upper-bounded by $2M$ in bits, whereas the information-theoretical capacity with ideal ADC increases with the transmit power logarithmically. Furthermore, due to the severe nonlinearity incurred, the use of low-resolution ADC has significant impact on various system designs, such as input signal optimization, channel estimation, etc., for which more research effort is needed to achieve the full potential of low-resolution ADC. Moreover, it is still unclear how a similar technique could be applied at the transmitter side to reduce the transmitting RF chain cost. 






\begin{figure}
\centering
\includegraphics[scale=0.7]{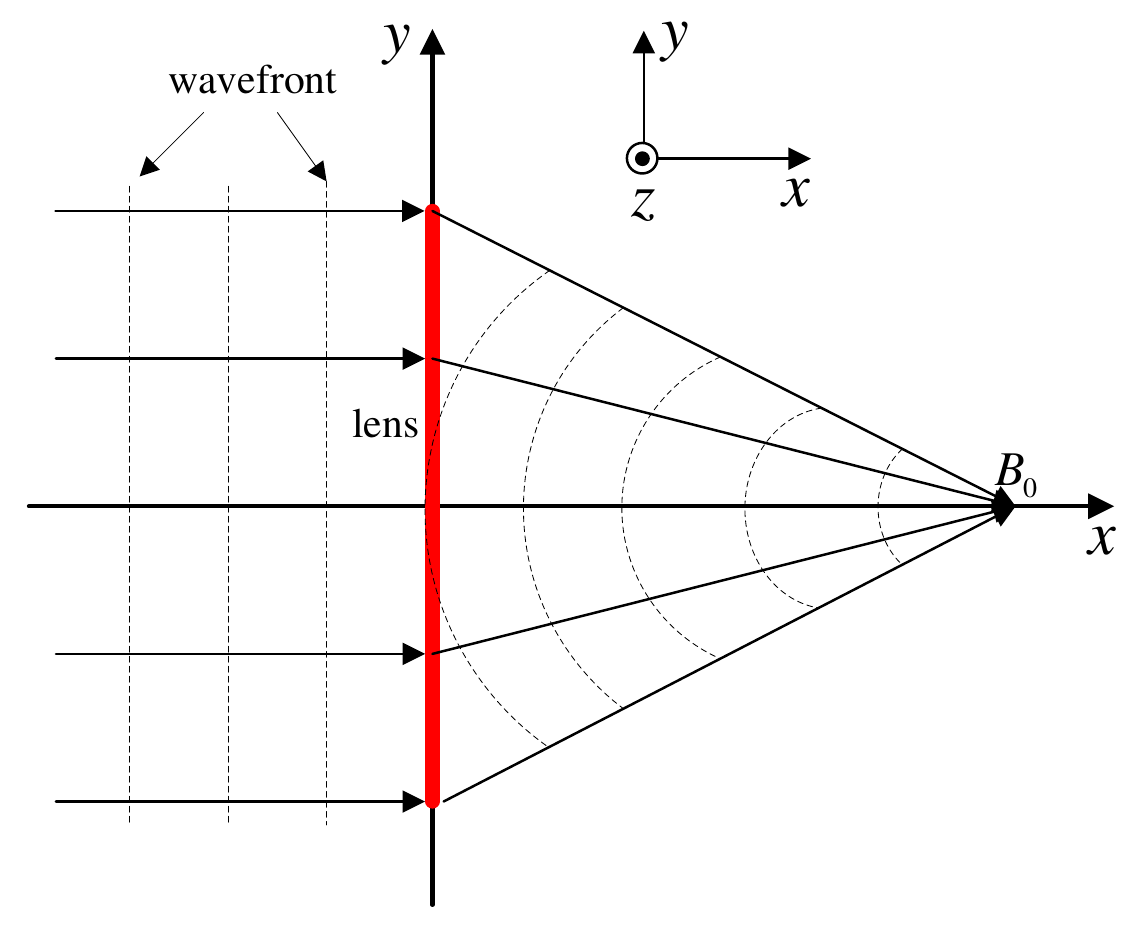}
\caption{The top view of a planar EM lens placed in the y-z plane with focal point $B_0$ for normal incident plane wave.}\label{F:EMlensTopView}
\end{figure}

\section{MmWave MIMO with Lens Antenna Arrays}
In this section, we present an alternative technique to achieve cost-effective mmWave communications by utilizing the advanced lens antenna arrays \cite{553,485,823,627}, which in general consist of an electromagnetic (EM) lens with energy focusing capability and a matching antenna array with elements located in the focal region of the lens.
\subsection{Electromagnetic Lens: Principle of Operation}
Similar to optical lenses, an EM lens is a transmissive device that is able to alter the propagation directions of the EM rays to achieve energy focusing or beam collimation. 
    EM lenses can be commonly implemented with three different techniques \cite{823}: i) the dielectric lenses (e.g., convex lenses) made of dielectric materials with carefully designed front and/or rear surfaces; ii) the traditional planar lenses consisting of arrays of transmitting and receiving antennas connected via transmission lines with different lengths; and iii) the modern compact planar lenses composed of sub-wavelength spatial phase shifters with periodic inductive and capacitive structures. Regardless of the actual implementation methods, the fundamental principle of EM lenses is to provide different phase shifting (or delays) for EM rays arriving at different points on the lens aperture, so as to achieve constructive superposition at the desired locations (focal region) and destructive cancelation otherwise. As a simple illustration, Fig.~\ref{F:EMlensTopView} shows the top view of a planar EM lens with negligible thickness placed in the y-z plane. To achieve energy focusing at point $B_0$ for normal incident plane waves, it is intuitive that the EM rays arriving at  the center of the lens should incur a larger phase shift than those at the edges, since after passing through the lens, they need to travel a shorter distance before arriving at $B_0$. Interestingly, with such a phase shifting profile of the lens designed only for normal incident waves, the resulting EM lens is able to achieve similar energy focusing for oblique incident plane waves with arbitrary arriving angles, but at different energy focusing locations \cite{823}. Such an angle-dependent energy focusing capability of EM lenses offers a unique opportunity to design lens antenna arrays, as discussed next.



\begin{figure}
\centering
\includegraphics[scale=0.6]{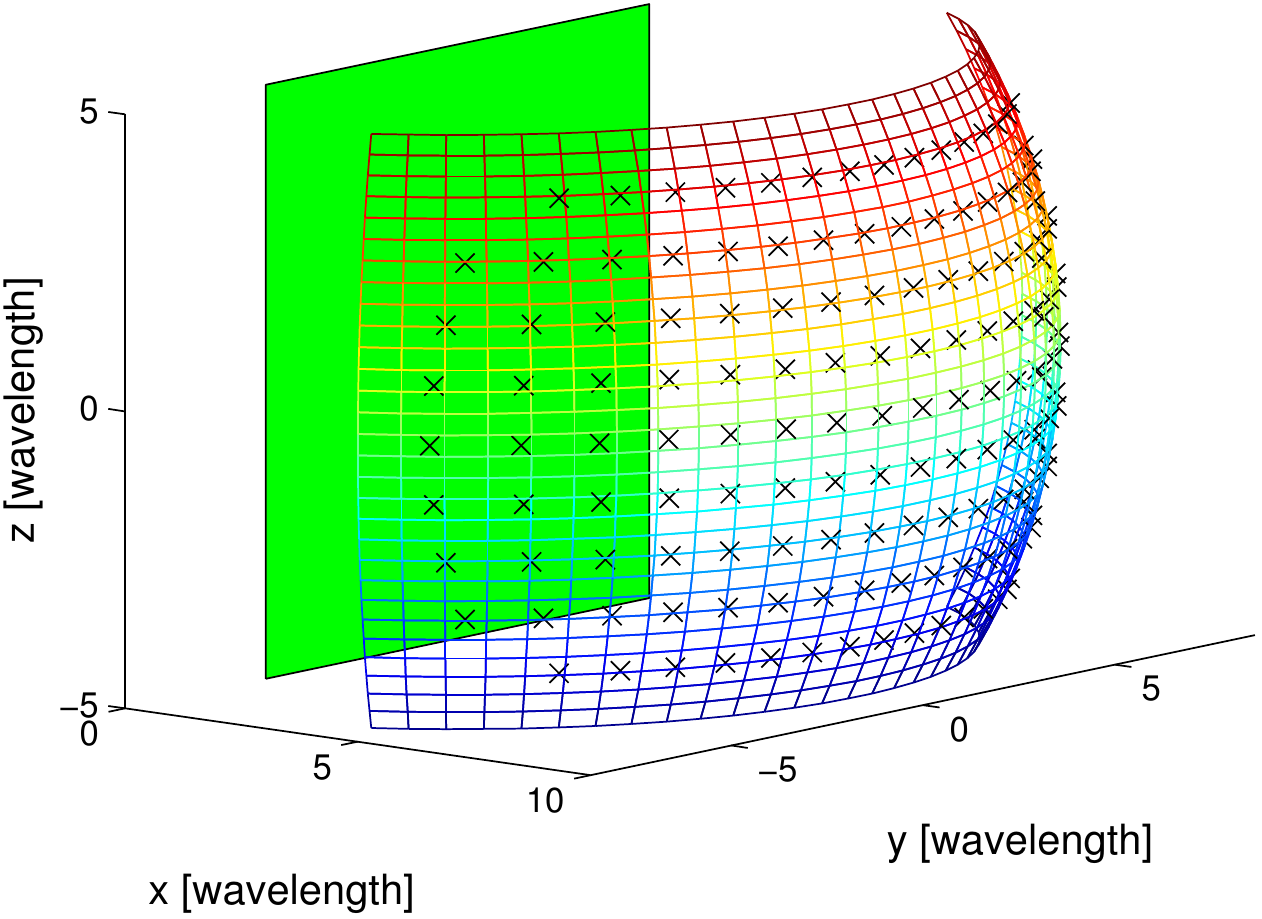} \qquad \includegraphics[scale=0.3]{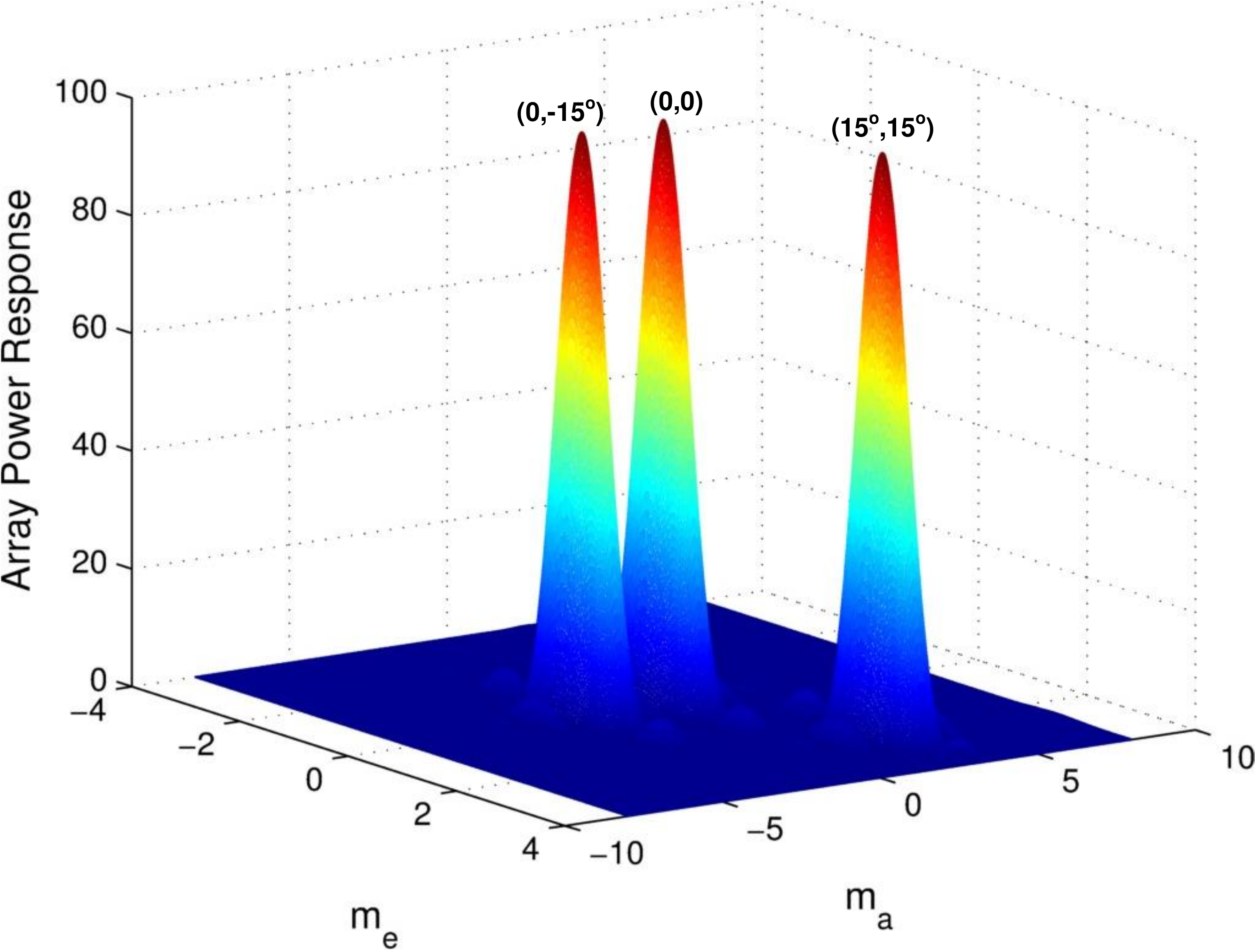} 
\\
  (a)  \hspace{7cm} (b)
\caption{(a) An example of full-dimensional lens antenna array; (b) The corresponding power response for three different signal directions $(\theta, \phi)=(0,0), (0, -15^\circ), (15^\circ, 15^\circ)$, where $m_e$ and $m_a$ denote the elevation and azimuth indices of each antenna element.}\label{F:FDLensAndPowerResponse}
\end{figure}

\subsection{Lens Antenna Array}\label{sec:lensArray}
Fig.~\ref{F:FDLensAndPowerResponse}(a) illustrates an example lens antenna array in three-dimensional (3D) coordinate system. We assume that a planar EM lens of electric dimension $\D_y \times \D_z$ (i.e., the lens' physical dimension normalized by wavelength)  is placed in the y-z plane and centered at the origin. The elements of the matching antenna array are placed in the focal surface of the EM lens, which is a hemisphere surface around the lens' center with a certain radius or focal length. Therefore, the antenna position of the $m$th array element relative to the lens center can be parameterized by its elevation and azimuth angles $\theta_m\in [-\Theta/2, \Theta/2]$ and $\phi_m \in [-\Phi/2, \Phi/2]$,  where $\Theta$ and $\Phi$ respectively denote the elevation and azimuth  angles to be covered by the lens antenna array. 
 We assume that the antennas are placed  such that $\sin\theta_m=m_e/\D_z$, where $m_e=0,\pm 1, \cdots, \pm \lfloor \D_z \sin (\Theta/2)\rfloor$ denotes the elevation index of antenna $m$, with $\lfloor x \rfloor$ denoting the largest integer no greater than $x$. Similarly, $\sin \phi_m = m_a/(\D_y \cos\theta_m)$, where $m_a=0,\pm 1, \cdots, \pm \lfloor \D_y \cos \theta_m \sin(\Phi/2)\rfloor$ denotes the azimuth index of antenna $m$. In other words, each antenna $m$ of the lens antenna array is parameterized by both its elevation and azimuth indices $(m_e, m_a)$.

With similar derivation as in \cite{823}, the array response of the lens antenna array as a function of the signal's elevation and azimuth angles $(\theta, \phi)$ can be expressed as
\begin{align}
a_m(\theta, \phi) \approx \sqrt{\D_y \D_z} \sinc (m_e - \D_z \sin \theta) \sinc (m_a - \D_y \cos \theta \sin \phi), \notag 
\end{align}
where $\sinc(x)\triangleq \sin(\pi x)/(\pi x)$ is the ``sinc'' function. The ``sinc-'' type array response shows that for any incident/departure signal from/to a particular direction $(\theta, \phi)$, only those antennas located in the close vicinity of the focal point could receive/steer significant power; whereas the power of all other antennas located far away is almost negligible. As a result, any two simultaneously received/transmitted signals with sufficiently separated directions can be effectively discriminated over different antenna elements. This is illustrated by Fig.~\ref{F:FDLensAndPowerResponse}(b) which shows the power responses of the lens antenna array for three different signal directions. 
It is observed that the lens antenna array is able to separate signals both in elevation and azimuth directions, where their resolutions can be enhanced by increasing $\D_z$ and $\D_y$, respectively. This general lens antenna array design is termed {\it full-dimensional (FD) lens array}. 


\subsection{Prototype and Measurement Results}
\begin{figure}
\centering
\includegraphics[scale=0.04]{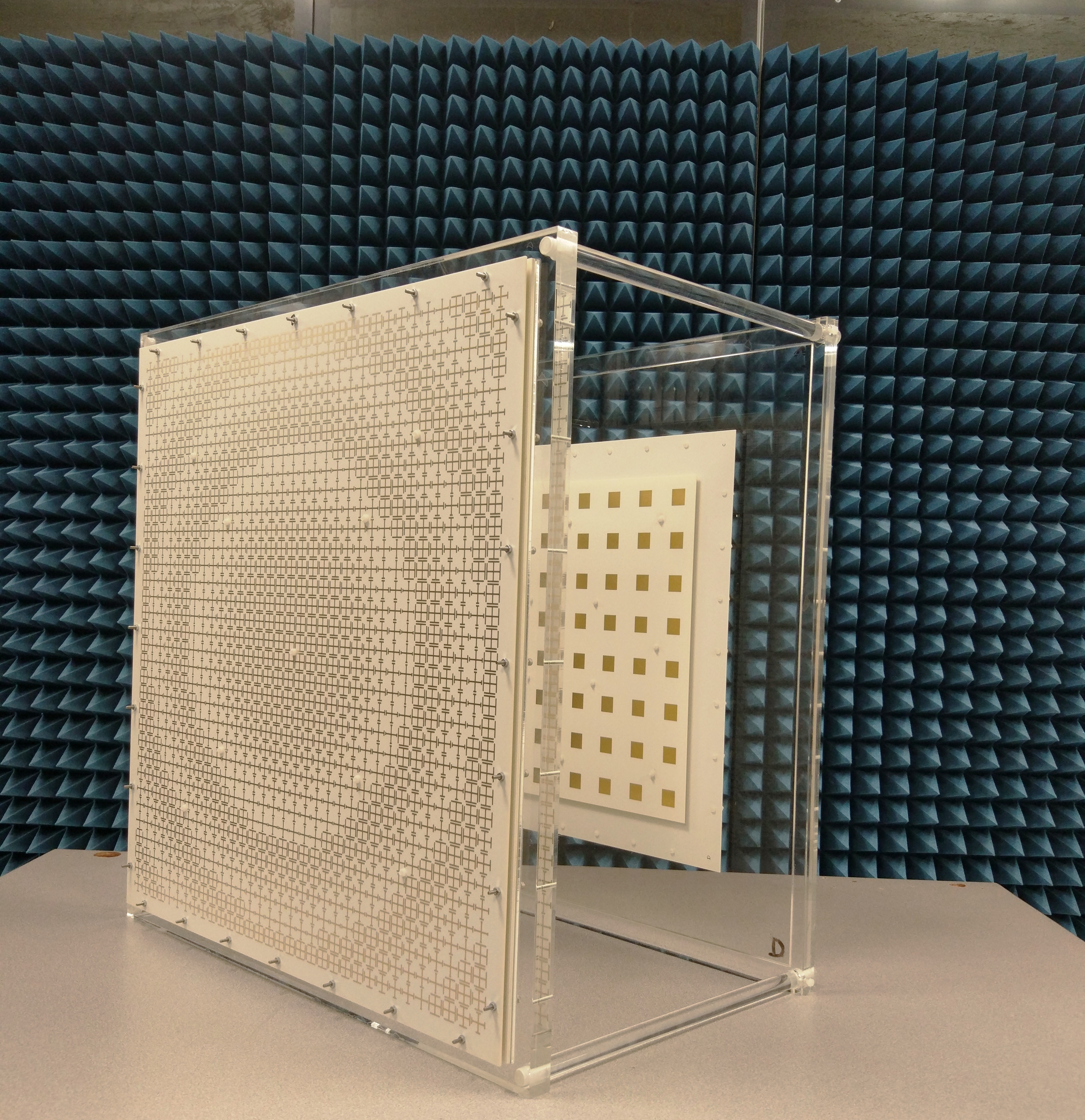}
\includegraphics[scale=0.5]{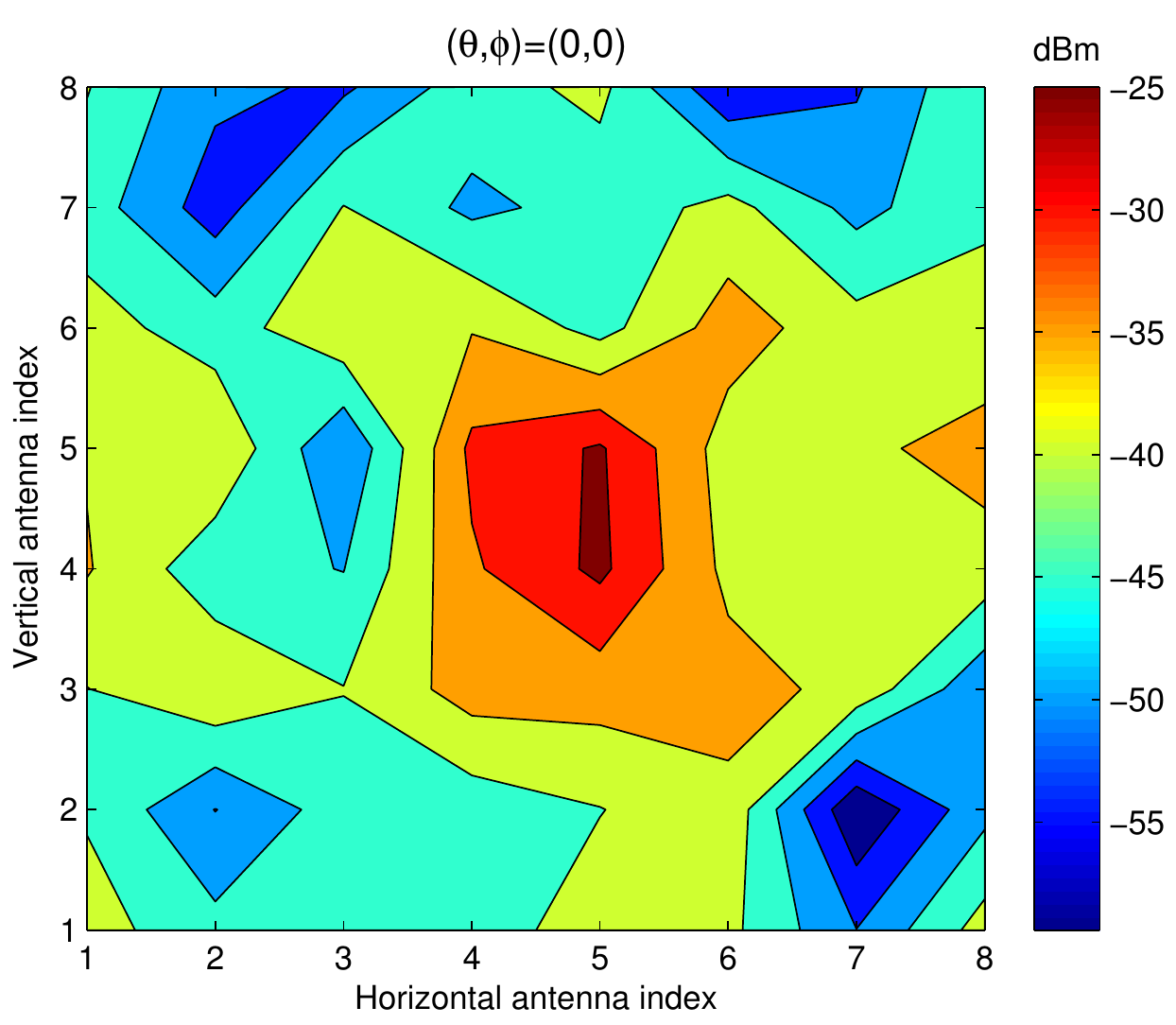} \includegraphics[scale=0.5]{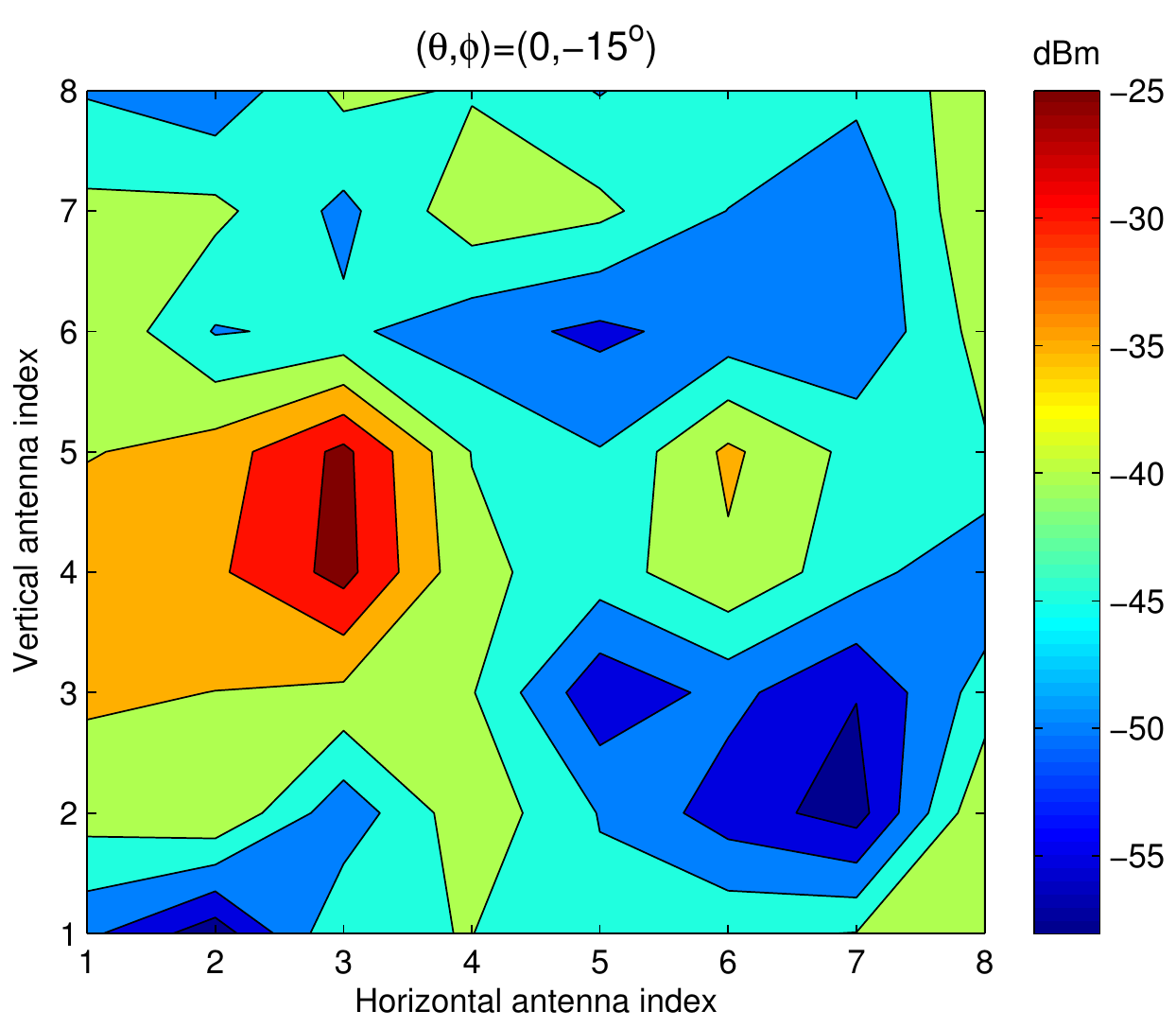}
\caption{The lens antenna array designed by the National University of Singapore and its contour plot of the measured received power for two different signal directions.}\label{F:MeasuredPower}
\end{figure}

To practically validate the energy focusing property of lens antenna arrays, a prototyping lens array has been designed, fabricated and tested, which is shown in Fig.~\ref{F:MeasuredPower}.  For ease of practical fabrication, the central frequency is chosen to be 5.8GHz and the array elements are uniformly placed on the antenna plane that is in parallel with the lens plane, rather than on the focal surface that is however expected to yield better energy focusing performance theoretically. The design of lens arrays for higher frequencies and with array elements placed on the focal surface will be left as future work. Note that theoretically, the energy focusing property of lens arrays to be verified here is irrespective of the actual operating frequency. 
 The designed planar EM lens has the physical dimension of 52.8cm $\times$ 52.8cm ($\D_z=\D_y\approx 10$), and an $8\times 8$ patch antenna array with adjacent elements separated by 4.2cm is placed on the antenna plane that is 25cm away from the lens plane, with each array element having two ports for horizontally and vertically polarized waves, respectively. 

 The designed lens antenna array was tested in a chamber environment as a receiving array, with the transmitter being a horn antenna with gain 10dBi and transmitting power $10$dBm. The transmitter and receiver are separated by $5$m. Based on the measurement results, Fig.~\ref{F:MeasuredPower} shows the contour plot of the received signal power by different elements of the lens array for signals arriving at $(\theta, \phi)=(0,0)$ and $(0, -15^\circ)$, respectively. It is noted that for normal incident wave, those antennas with indices around $(5,4)$ receive significantly higher power than other elements, which verifies the energy focusing capability of the designed lens array. Furthermore,  when the signal arrives at a different direction $(\theta, \phi)=(0, -15^\circ)$, the energy focusing antennas are shifted to those with indices around $(3,4)$, which experimentally verifies that the energy focusing antennas of lens arrays  indeed vary with the signal directions.

\subsection{MmWave Lens MIMO: Cost-Effective and Capacity-Optimal Single-Carrier Communication}\label{sec:mmWaveLensWithSC}
Lens antenna arrays are particularly suitable for mmWave communications for two main reasons. Firstly, at mmWave frequencies, the EM lens and also the resulting lens array have compact size due to the small signal wavelength. 
Secondly, 
the angle-dependent energy focusing capability of the lens array, together with the angular sparsity of mmWave channels, offers an appealing cost-effective solution to achieve capacity-optimal mmWave MIMO communications with the low-complexity antenna selection and SC modulation schemes, even for frequency-selective wide-band communications, as will be detailed as follows.

\begin{figure}
\centering
\includegraphics[scale=0.6]{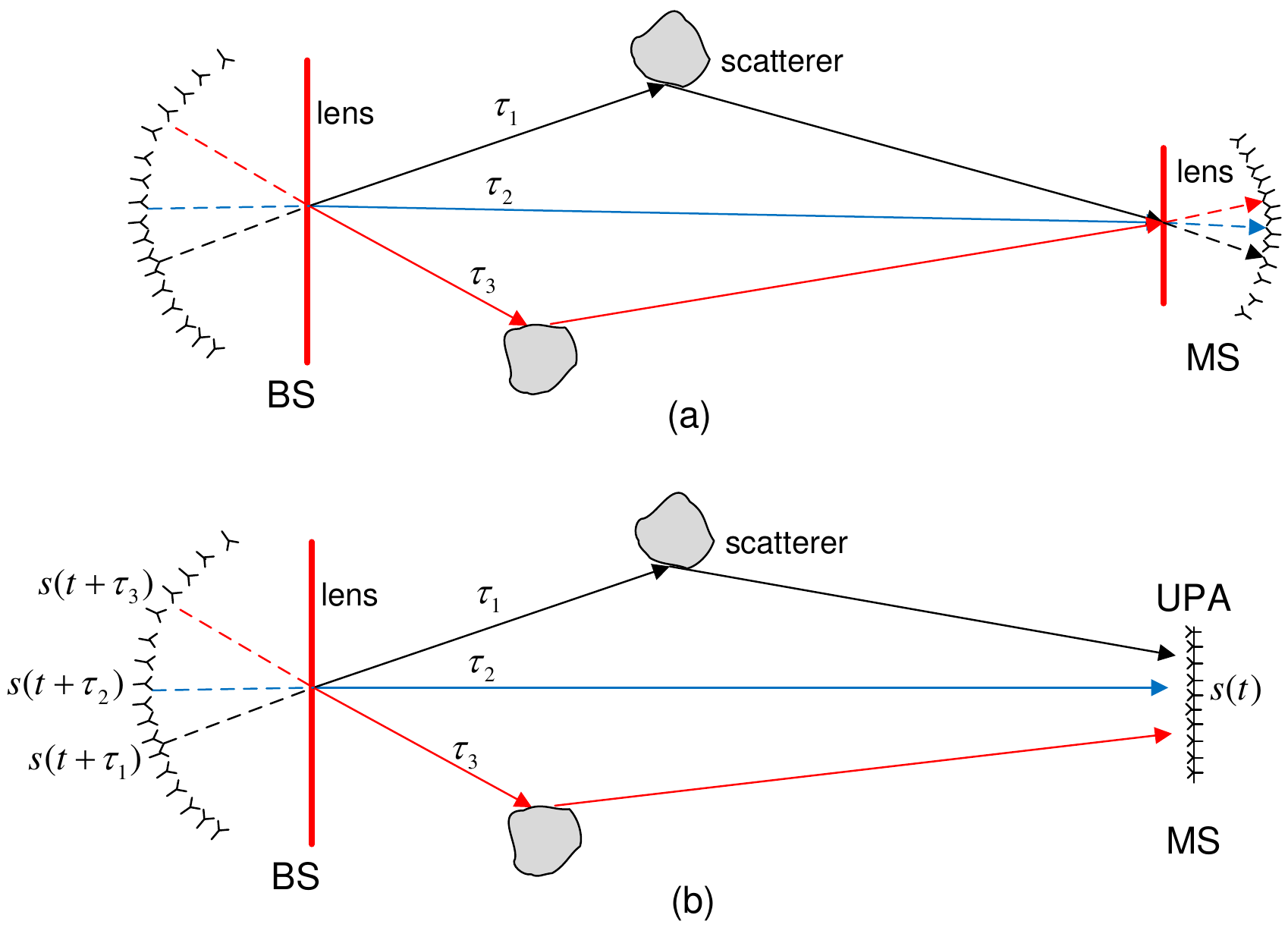}
\caption{MmWave lens MIMO system with two different setups: (a) Double-sided lens with large lens arrays at both BS and MS to completely decouple the multi-path signals; (b) Single-sided lens with large lens array at the BS and UPA at the MS, with the ISI mitigated by  path delay pre-compensation at the BS, where $\tau_l$ denotes the delay of the $l$th path.}\label{F:mmWaveLensMIMOMP}
\end{figure}

We first consider a point-to-point mmWave {\it double-sided} lens MIMO system, where both the BS and the MS are equipped with large lens antenna arrays with the number of antenna elements much larger than the number of multi-paths of the mmWave channel. Thus, both the BS and MS offer sufficiently fine angle resolutions to discriminate all the multi-path signals, i.e.,  the signals to (from) the distinct channel paths are steered (focused) by (on) {\it non-overlapping} antenna subsets at the transmitter (receiver),  as illustrated in Fig.~\ref{F:mmWaveLensMIMOMP}(a). By selecting the corresponding energy-focusing antennas, the large lens MIMO system with $L$ multi-paths is essentially decoupled into $L$ parallel small-MIMO channels, each corresponding to only one of the multi-path and hence is always frequency-nonselective (flat), regardless of the signal bandwidth. This is appealing for two reasons. Firstly, the number of antennas required to be activate, and hence the number of RF chains to be equipped, only depends on the number of multi-paths $L$, rather than that of the actual antenna elements of the lens arrays. Secondly, with the multi-path signals decoupled over different array elements, the detrimental ISI effect in wide-band mmWave communications  is inherently resolved without the need of sophisticated ISI mitigation techniques such as OFDM or equalization. As a result, the channel capacity is simply achieved by the low-complexity SC communications, by multiplexing $L$ data streams each over one of the multi-paths with parallel small-MIMO processing. We term such a MIMO spatial multiplexing scheme enabled by mmWave lens MIMO as {\it path division multiplexing} (PDM) \cite{823}.

For some practical scenarios where the lens array at the MS is small and hence has insufficient angle resolution to resolve all the multi-paths, or for {\it single-sided} lens MIMO system with the MS equipped with the conventional antenna array (such as UPA) as shown in Fig.~\ref{F:mmWaveLensMIMOMP}(b), SC transmission is still promising, as long as the multi-paths are well separated by the BS lens array. This is because the ISI can be eliminated by path delay pre-compensation at the BS, which is feasible since each of its antenna element corresponds to at most one channel path, and hence only one delay value needs to be pre-compensated for, as illustrated in Fig.~\ref{F:mmWaveLensMIMOMP}(b) for the example of downlink communication with three paths. With perfect path delay pre-compensation, all channel paths effectively have identical delays (arriving time) at the receiver, which leads to frequency-nonselective channels regardless of the signal bandwidth. This is in a sharp contrast to MIMO systems with conventional arrays, where each element receives/transmits signals from/to all channel paths each with a different path delay in general, which renders the simple path delay pre-compensation technique for ISI mitigation infeasible, and thus more sophisticated techniques like OFDM or equalization need to be used.

\section{Performance Comparison}
In this section, numerical results are provided to compare the performance of the lens-based mmWave system with that based on the conventional UPA.  We consider a point-to-point mmWave MIMO system at $28$GHz. We assume that the MS is equipped with a small UPA of four elements, whereas the BS either has a FD lens antenna array (proposed), or a UPA (benchmark) with adjacent elements separated by half wavelength. We assume that both the lens array and the UPA at the BS have the same effective aperture $\D_y \times \D_z=10\times 10$, and are designed to have the same azimuth and elevation coverage angles $\Phi=120^\circ$ and $\Theta=60^\circ$, respectively. Thus, the total number of BS antennas for UPA is $M_{\upa}=400$ and that for the lens array is $M_{\lens}=149$. We assume that the mmWave channel has three paths, with the azimuth and elevation signal angles uniformly distributed in $[-60^\circ, 60^\circ]$ and $[-30^\circ, 30^\circ]$, respectively, and the path delays uniformly distributed in $[0, T_m]$, with $T_m=100$ns being the maximum path delay. Furthermore, the path loss and the power division among different paths are generated based on the model developed in \cite{565}. We assume that the available bandwidth is $B=500$MHz. As a result, we have $\mu\triangleq BT_m=50\gg 1$, so that the system is wide-band and frequency-selective in general. For the UPA-based system, MIMO-OFDM transmission is assumed, with $N=512$ sub-carriers and $\mu=50$ cyclic prefix (CP) symbols. On the other hand, for the lens array-based system, we employ the low-complexity SC transmission with path delay pre-compensation (see Section~\ref{sec:mmWaveLensWithSC}).

\subsection{Rate Comparison}
\begin{figure}
\centering
\includegraphics[scale=0.7]{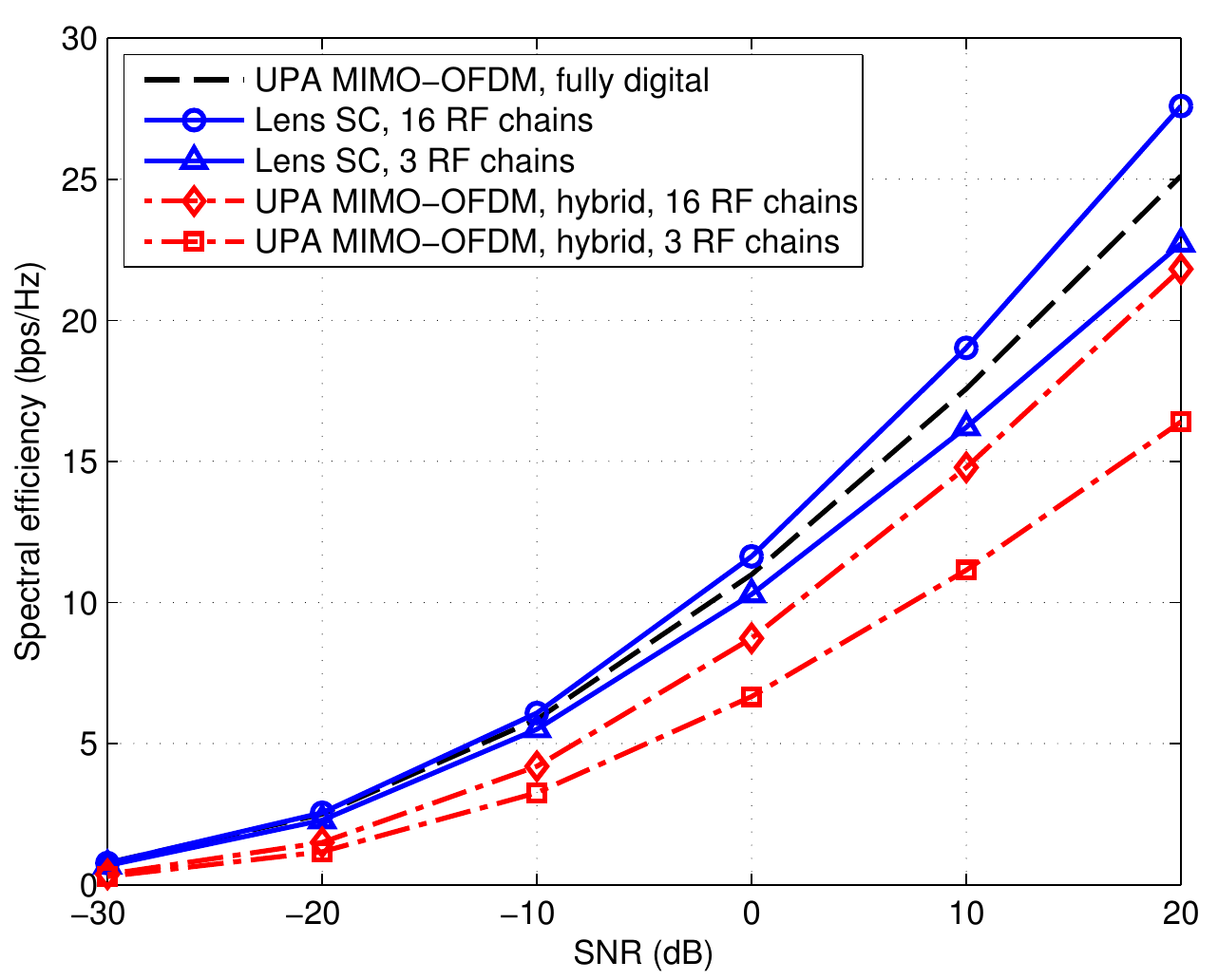}
\caption{The spectrum efficiency comparison of lens-based SC transmission versus the UPA-based MIMO-OFDM transmission with hybrid precoding.}\label{F:CompareRateVSSNR}
\end{figure}

Fig.~\ref{F:CompareRateVSSNR} shows the spectrum efficiency versus the average SNR of each receive UPA antenna for the two antenna systems with $M_{\rf}=3$ and $16$ RF chains at the BS, respectively. Note that since the number of antennas at the MS is small, we assume that it has the same number of RF chains as the antenna elements and hence can apply the fully digital baseband processing. On the other hand, to cater for the limited number of RF chains at the BS side, the simple power based antenna selection scheme is adopted for the lens system, whereas the approximate Gram-Schmidt based hybrid precoding scheme for MIMO-OFDM proposed in \cite{825} is applied for the UPA system,  where each RF beamforming vector is chosen from the beamsteering codebook of size $256$ which is obtained by uniformly quantizing the azimuth and elevation angles. As a performance upper bound for the hybrid precoding scheme, we also consider the fully digital UPA-based MIMO-OFDM scheme with the optimal eigenmode transmission and water-filling power allocation, by assuming that the BS has full RF chains, i.e., $M_{\mathrm{rf}}=400$. It is observed that even with $M_\rf=3$, the lens array based SC transmission scheme already achieves very close performance as the fully digital UPA-based MIMO-OFDM, whereas the hybrid precoding scheme performs significantly worse than the fully digital scheme. As $M_\rf$ increases to $16$, the lens system even outperforms the fully digital MIMO-OFDM, which is expected since the proposed lens-based SC transmission requires no CP overhead as that in MIMO-OFDM. On the other hand, the hybrid precoding scheme with $M_\rf=16$ still incurs a notable performance loss compared to the fully digital scheme. 


\subsection{Power Consumption Comparison}
Next, we compare the BS power consumption for the various schemes shown in Fig.~\ref{F:CompareRateVSSNR}. For the benchmark fully digital UPA system, the major power consumption is due to the various components of the RF chains. Since one RF chain is needed for each antenna element for fully digital implementation, we have $P_{\upa,\dig}=M_\upa P_{\rf}$, with $P_{\rf}$ denoting the power consumption of each RF chain. On the other hand, for UPA-based hybrid precoding scheme, though the total power consumption of RF chains  can be reduced by using fewer RF chains, it incurs an additional power consumption cost associated with the phase shifters, which requires a total number of $M_\upa M_\rf$ for the full-array hybrid architecture in \cite{825}. Thus, the main power consumption for the UPA-based hybrid precoding scheme can be modeled as $P_{\upa,\hybrid}=M_\rf P_{\rf}+M_\upa M_\rf P_{\mathrm{ps}}$, with $P_{\mathrm{ps}}$ denoting the power consumption required for each phase shifter. By replacing the phase shifters with the low-complexity analog switches, the power consumption for the lens-based antenna selection scheme can be similarly modeled as $P_{\lens}=M_\rf P_\rf + M_\lens M_\rf P_{\mathrm{sw}}$, with $P_{\mathrm{sw}}$ denoting the power consumption for each analog switch. Note that in the above models, we have ignored the power consumption associated with the signal radiation and baseband signal processing, for which the lens-based system should consume the least power due to the simple SC transmission. By setting $P_{\rf}=250$mW, $P_{\mathrm{sw}}=5$mW, and $P_{\mathrm{ps}}=15$mW \cite{824}, Table~\ref{table:powerCompare} summarizes the power consumption for the three schemes in Fig.~\ref{F:CompareRateVSSNR}. It is shown that due to the large number of RF chains required, the fully digital precoding consumes a substantial power of $100$W. With hybrid precoding scheme, the power consumption reduces to $18.75$W if $M_\rf=3$ RF chains are used, but it increases to $100$W if $M_\rf$ increases to $16$. This is attributed to the excessively large number of phase shifters ($400\times 16$) needed for the  hybrid precoding  as well as the considerable power consumption for each phase shifter. In contrast, for the lens MIMO system, the power consumption dramatically reduces to $2.985$W for $M_\rf=3$ and $15.92$W for $M_\rf=16$, thanks to the reduced number of antenna elements needed for lens array as well as the use of less power-consuming switches compared to phase shifters in the hybrid precoding scheme.


\begin{table}
\centering
\caption{Power consumption comparison for the various schemes.}\label{table:powerCompare}
\begin{tabular}{|p{4cm}|p{2cm}|p{2cm}|}
\hline
&  \multicolumn{2}{c|}{\bf Power consumption (W)} \\
\hline
UPA fully digital  & \multicolumn{2}{c|}{100}\\
\hline
\hline
 & $M_\rf=3$ & $M_\rf=16$ \\
\hline
UPA, hybrid precoding &  18.75 & 100\\
\hline
Lens, antenna selection &  2.985 & 15.92\\
\hline
\end{tabular}
\end{table}


%
%
%

\section{Extensions and Future Work}
There are many directions along which  mmWave lens MIMO systems can be further investigated. For example, the energy focusing property of lens antenna arrays also offers a unique opportunity for implementing efficient channel estimations with low training and feedback overhead. With limited number of  RF chains available, the energy focusing antennas can be searched via sequential training with low-complexity receiver energy detection and comparison \cite{830}, which significantly reduces the size of the MIMO channel to be estimated for data transmission. Another important problem is to investigate the performance and size trade-off for mmWave lens MIMO systems. Notice that compared to the traditional UPAs, lens antenna array requires  a sufficient gap between the EM lens and antenna array, which increases the transceiver size but is required to achieve good energy-focusing performance of the lens array. Last but not least, mmWave lens MIMO systems in multi-user setups require further studies. In this case, the PDM concept could be extended to path division multiple access (PDMA) for simultaneously serving multiple users via different multi-paths. In particular, for the practical case of one-sided lens MIMO system where the multi-path signals of different users could be effectively separated at the BS, optimal beamforming schemes at the MSs still need to be designed to enhance the performance when only limited RF chains are available.

\section{Conclusions}
In this article, we first review the existing techniques that offer different trade-offs between performance and RF chain cost for mmWave communication systems, and then focus our investigation on the mmWave lens MIMO technique, which is shown to be able to achieve the capacity-optimal performance, and also significantly reduce the signal processing complexity as well as hardware and power consumption costs. The performance gains in both spectrum efficiency and energy efficiency are also shown as compared to other existing benchmark designs.

\section*{Acknowledgement}
The authors would like to thank Zhi Ning Chen and Shunli Li in the National University of Singapore for helpful discussions as well as the help to produce Fig.~\ref{F:MeasuredPower}.

\bibliographystyle{IEEEtran}
\bibliography{IEEEabrv,IEEEfull}

\end{document}